\documentclass[sn-nature]{sn-jnl}

\usepackage{graphicx}%
\usepackage{multirow}%
\usepackage{amsmath,amssymb,amsfonts}%
\usepackage{amsthm}%
\usepackage{mathrsfs}%
\usepackage[title]{appendix}%
\usepackage{xcolor}%
\usepackage{textcomp}%
\usepackage{manyfoot}%
\usepackage{booktabs}%
\usepackage{algorithm}%
\usepackage{algorithmicx}%
\usepackage{algpseudocode}%
\usepackage{listings}%
\usepackage{gensymb}
\usepackage{comment}
\usepackage{bm}
\usepackage{lineno}

\renewcommand{\emph}[1]{\textit{#1}}

\usepackage{soul}

\usepackage{pdfpages}
\usepackage{float}

\theoremstyle{thmstyleone}%
\theoremstyle{thmstyletwo}%

\theoremstyle{thmstylethree}%

\begin{document}

\title{Probing nuclear interactions à la Rutherford: Insights on $^4$He from $\alpha$ scattering}


\author[1,2]{\fnm{F.} \sur{Cappuzzello}} 
\author*[1,2]{\fnm{V.} \sur{Soukeras}} \email{vasileios.soukeras@lns.infn.it}
\author[3,4]{\fnm{S.} \sur{Bacca}} 
\author[2]{\fnm{D.} \sur{Carbone}} 
\author[2]{\fnm{M.} \sur{Cavallaro}} 
\author[5]{\fnm{L. C.} \sur{Chamon}} 
\author[1,6]{\fnm{I.} \sur{Lombardo}} 
\author[7,8]{\fnm{G.} \sur{Orlandini}} 
\author[9]{\fnm{M.} \sur{Viviani}}
\author[2]{\fnm{C.} \sur{Agodi}} 
\author[10]{\fnm{H -W.} \sur{Becker}} 
\author[1,2]{\fnm{G. A.} \sur{Brischetto}} 
\author[2]{\fnm{S.} \sur{Calabrese}} 
\author[11]{\fnm{C.} \sur{Ciampi}}
\author[12,13]{\fnm{M.} \sur{Cicerchia}} 
\author[14]{\fnm{M.} \sur{Cinausero}} 
\author[1,2]{\fnm{I.} \sur{Ciraldo}} 
\author[15,16]{\fnm{D.} \sur{Dell’Aquila}} 
\author[2]{\fnm{M.} \sur{Fisichella}} 
\author[11]{\fnm{C.} \sur{Frosin}} 
\author[2,17]{\fnm{A.} \sur{Hacisalihoglu}} 
\author[18]{\fnm{M.} \sur{Hilcker}} 
\author[8]{\fnm{A.} \sur{Kievsky}}
\author[19,20]{\fnm{Y.} \sur{Kucuk}} 
\author[12]{\fnm{T.} \sur{Marchi}}  
\author[1,2]{\fnm{O.} \sur{Sgouros}} 
\author[1,2]{\fnm{A.} \sur{Spatafora}} 
\author[2]{\fnm{D.} \sur{Torresi}} 
\author[15,16]{\fnm{M.} \sur{Vigilante}} 
\author[20]{\fnm{A.} \sur{Yildirim}}

\affil[1]{\orgdiv{Dipartimento di Fisica e Astronomia “Ettore Majorana”}, \orgname{Università di Catania}, \orgaddress{\city{Catania}, \country{Italy}}}

\affil[2]{\orgname{INFN – Laboratori Nazionali del Sud (LNS)}, \orgaddress{\city{Catania}, \country{Italy}}}

\affil[3]{\orgdiv{Institut für Kernphysik}, \orgname{Johannes Gutenberg-Universität Mainz}, \orgaddress{\city{Mainz}, \country{Germany}}}

\affil[4]{\orgdiv{Helmholtz-Institut Mainz}, \orgname{Johannes Gutenberg-Universität Mainz}, \orgaddress{\city{Mainz}, \country{Germany}}}

\affil[5]{\orgdiv{Departamento de Física Nuclear}, \orgname{Instituto de Física da Universidade de São Paulo}, \orgaddress{\city{São Paulo}, \country{Brazil}}}

\affil[6]{\orgname{INFN – Sezione di Catania}, \orgaddress{\city{Catania}, \country{Italy}}}

\affil[7]{\orgdiv{Dipartimento di Fisica}, \orgname{Università di Trento}, \orgaddress{\city{Trento}, \country{Italy}}}

\affil[8]{\orgname{INFN - Trento Institute for Fundamental Physics and Applications (TIFPA)}, \orgaddress{\city{Trento}, \country{Italy}}}

\affil[9]{\orgname{INFN – Sezione di Pisa}, \orgaddress{ \city{Pisa}, \country{Italy}}}

\affil[10]{\orgname{Ruhr-Universitat Bochum}, \orgaddress{\city{Bochum}, \country{Germany}}}

\affil[11]{\orgname{INFN – Sezione di Firenze}, \orgaddress{\city{Firenze}, \country{Italy}}}

\affil[12]{\orgname{Dipartimento di Fisica e Astronomia, Università di Padova}, \orgaddress{\city{Padova}, \country{Italy}}}

\affil[13]{\orgname{INFN – Sezione di Padova}, \orgaddress{\city{Padova}, \country{Italy}}}

\affil[14]{\orgname{INFN – Laboratori Nazionali di Legnaro (LNL)}, \orgaddress{\city{Legnaro}, \country{Italy}}}

\affil[15]{\orgname{Dipartimento di Fisica "Ettore Pancini", Università degli Studi di Napoli “Federico II”}, \orgaddress{\city{Napoli}, \country{Italy}}}

\affil[16]{\orgname{INFN – Sezione di Napoli}, \orgaddress{\city{Napoli}, \country{Italy}}}

\affil[17]{\orgname{Institute of Natural Science, Karadeniz Teknik Universitesi}, \orgaddress{\city{Trabzon}, \country{Turkey}}}

\affil[18]{\orgdiv{Department of Physics, Institute for Nuclear Physics}, \orgname{Technische Universität Darmstadt}, \orgaddress{\city{Darmstadt}, \country{Germany}}}

\affil[19]{\orgname{Turkish Accelerator and Radiation Laboratory (TARLA)}, \orgaddress{\city{Gölbaşı/Ankara}, \country{Turkiye}}}

\affil[20]{\orgname{Akdeniz University}, \orgaddress{\city{Antalya}, \country{Turkey}}}



\abstract{Nuclear interactions play a key role for the stability of atomic nuclei and stellar environments. Successful parametrization and models of these interactions, developed in the last decades, accurately reproduce all the proton and neutron scattering data, besides the properties of few-body nuclear systems.
However, recent electron scattering results focusing on the first excited resonant state of $^4$He nucleus, reveal a puzzling situation suggesting potential gaps in our understanding of the nuclear phenomenology.
Here, we report a new study of such $^4$He resonance by $^4$He + $^4$He scattering featuring data of unprecedented sensitivity and state-of-art analyses of the spectral line shape together with a phenomenological reaction modeling that incorporates the same nuclear densities employed in electron-scattering studies. Our analysis of the full set of experimental observables yields a reasonable description within the framework of current nuclear-interaction physics, thereby highlighting the need for further advancing the modeling of few-body open quantum systems.}

\maketitle

\section{Introduction}\label{sec1}
Ernest Rutherford's scattering experiments with $\alpha$ particles, aka $^4$He nuclei in their ground state, led to the groundbreaking discovery of the atomic nucleus over a century ago~\cite{Rutherford}, 
revolutionized our understanding of atomic structure and laid the foundation for modern nuclear physics.
Building on this rich legacy,
in this paper we delve into the contemporary exploration of $\alpha$ scattering employing new technologies and methods developed since Rutherford's time  with the goal of shedding light onto an intriguing quantum mechanical problem involving the $\alpha$ particle itself.

The $^4$He nucleus, is one of the simplest and most stable nuclei, constituted by two protons and two neutrons. Its ground state features a spin-parity $J^{\pi}$ = 0$^+$ and it is commonly indicated as the 0$^+_1$ state. Since its discovery~\cite{edinburg}, $^4$He has captivated the scientific community as it represents the most stable nuclear system, which makes it the seed of clustering in light- to medium-mass nuclei \cite{fre2010sch5} and of $\alpha$-decay radioactivity in heavy nuclei---the very same used by Rutherford. $^4$He also plays a crucial role in astrophysics as a fundamental component of stellar matter and a vital element in stellar nucleosynthesis~\cite{ili2007}.

The modeling of $^4$He in terms of 4 nucleons requires the solution of a quantum mechanical 4-body problem, where  nuclear interactions are a key element. The  task  of solving for the ground-state could be achieved already in 2001~\cite{kam2001prc64} by several so-called ab-initio methods~\cite{eks2023frp11}.
Progress accomplished in the past two decades has been  in increasing the precision  of nuclear interactions (referred to as nuclear Hamiltonians), either in a phenomenological way~\cite{pud1995prl74,wir1995prc51}, or by introducing the more systematic effective field theory (EFT) approach, particularly in the variant based on chiral symmetry~\cite{epe2009rmp91,ham2020rmp92}. Today, ground-state properties of $^4$He, such as binding energies and radii, are correctly described at about the percent level~\cite{mar2021prc103,kie2008jpg35}, and even nuclei as heavy as $^{208}$Pb~\cite{hu2022nat18} can be investigated by ab-initio nuclear theory.
However, the description of the $^4$He first excited state, a resonant state also featuring a spin-parity $J^{\pi}$ = 0$^+$ and therefore commonly indicated as the 0$^+_2$ state, still represents a challenge. This state is, in fact, situated between the $p$ + $^3$H  breakup channel starting at the proton-emission threshold $S_p$ = 19.813 MeV, and the $n$ + $^3$He  breakup channel  demarcated by the neutron emission threshold $S_n$ at 20.578 MeV.

The complication of computing such a state is intrinsic to its unbound resonant nature, as elucidated in a recent benchmark study \cite{due2025prc}. The position $E_r$--typically indicated as the centroid-energy above the $S_p$  threshold--and width $\mathit{\Gamma_r}$ were calculated with various theories 
\cite{due2025prc, hiy2004prc70,bac2013prl110,viv2020prc102,Aoy2016ptep2016,mic2023prl131,mei2023arx}, obtaining quite different results depending on the method used and on the implemented nuclear force.

\begin{figure}[!]
\begin{minipage}{1\textwidth}
\centering
\includegraphics[scale=0.50, trim = 0cm 0cm 0cm 17.5cm]{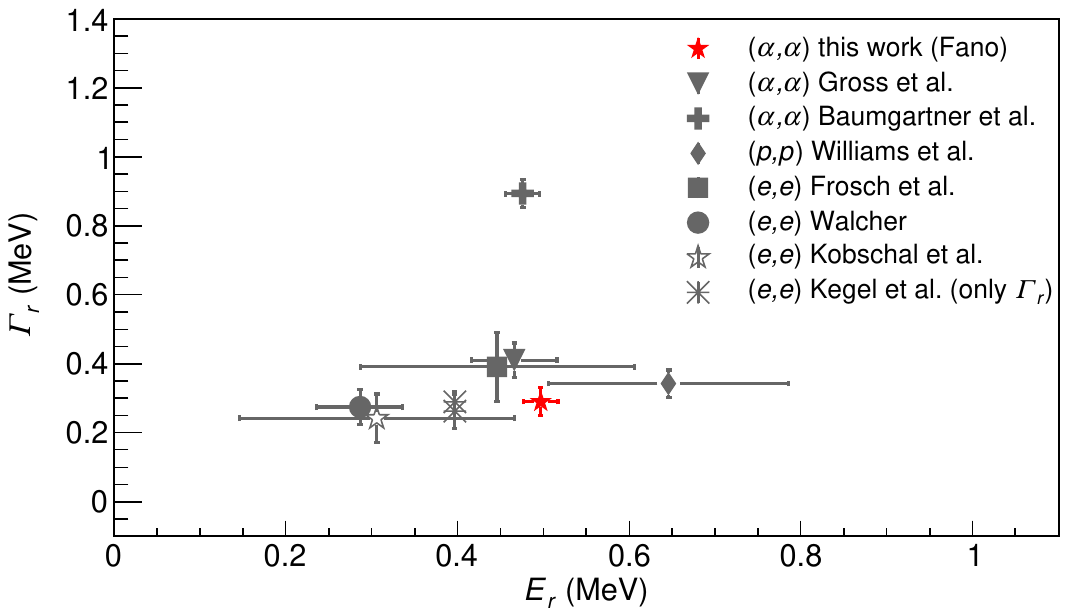}
\end{minipage}
\caption{ \textbf{Measured centroid-energy $E_r$ and width $\mathit{\Gamma_r}$ of the $0^+_1$ resonance~\cite{gro1969pr178,bau1981npa368,Wil1966,fro1968npa110,wal1970plb31,kob1983npa405,keg2023prl130}.} The $E_r$ is defined with respect to the proton-emission threshold $S_p$ = 19.813 MeV. The quoted error bars for the experimental datum of the present work (red) are discussed in Section \ref{Methods4}. The error bars of the other data points are given in Refs. \cite{gro1969pr178,bau1981npa368,Wil1966,fro1968npa110,wal1970plb31,kob1983npa405,keg2023prl130}.
\label{centroids_widths_b}}
\end{figure}

From an experimental perspective, determining the properties of the 0$^+_2$ state is also challenging.
 Inelastic scattering studies  on $^4$He targets using different projectiles, such as electrons~\cite{fro1968npa110,wal1970plb31,kob1983npa405, keg2023prl130}, protons~\cite{Wil1966} and $\alpha$-particles~\cite{gro1969pr178,bau1981npa368}, have in the past measured the position $E_r$ and width $\mathit{\Gamma_r}$ of this state. A large spread was observed, with a variation of about 0.6 and 0.7 MeV for $E_r$ and $\mathit{\Gamma_r}$, respectively, and with some data being inconsistent with the other (see Fig.~\ref{centroids_widths_b} and evaluation from Ref. \cite{til1992npa541}). Recently, the form factor $F_M$  of the transition  0$^+_1$ $\rightarrow$ 0$^+_2$, connected by an electric monopole operator, has attracted much attention in the nuclear physics community, see \cite{bac2013prl110, mei2023arx, mic2023prl131,viv2024fbs65} and Supplementary Methods (1.2). As a matter of fact, calculations present stark differences, and none of them reproduces all the three known properties ($E_r$, $\mathit{\Gamma_r}$ and $F_M$)  of the 0$^+_2$ state. This is an unsatisfactory  situation for what should be one of the best known nuclei. The properties of the first excited state of the $\alpha$ particle and their direct link to basic nuclear Hamiltonians are thus calling for further investigations.\par

Among the different probes which allow to populate the 0$^+_2$ state, inelastic scattering is the ideal toolbox as it requires minimal  rearrangement of the $^4$He nuclei. Nonetheless, notable differences arise depending on the projectiles employed in these studies. While electron scattering has the advantage of being perturbative, thus allowing for a direct connection of the cross section to the target properties, $\alpha$ scattering offers larger cross-sections, better state-selectivity and sensitivity to the density of the resonant state (see Supplementary Methods (1.1)), making it very effective for identifying resonant features, such as those of the 0$^+_2$ state in $^4$He. On the other hand, in the $\alpha$ - $\alpha$ scattering the many-body initial (ISI) and final state interactions (FSI) need to be accurately modeled to get detailed information of the involved states. Although this procedure can affect the overall uncertainty of the data analysis, modern nuclear reaction methods are able to describe the cross section with enough accuracy to disentangle different nuclear structure theories. In this framework, the resulting uncertainties can be significantly reduced by constraining the ISI and FSI with supplementary measurements of other observables such as elastic scattering and total reaction cross section.

The main challenge in the description of the 0$^+_2$ state is its location in the continuum, where other excited states may overlap and/or interfere.
Theoretical calculations~\cite{viv2020prc102} show that below  26 MeV of excitation energy (i.e., about 6 MeV above the  $S_p$ threshold) the relative angular momentum of the fragments is limited to $L$ = 0 ($S$-wave)  and $L$ = 1 ($P$-wave). 
Nonetheless, a number of different partial waves, namely the $^{1}S_0$, $^{3}P_0$, $^{3}P_1$, $^{3}P_2$, may contribute. Here, the molecular notation $^{2S+1}L_J$ is adopted to label the $^4$He states, where  $S$ is  the quantum number relative to the sum of the spins of the two fragments (namely $n$ and $^3$He or $p$ and $^3$H), and $J$($L$) the  total (orbital)  angular momentum  of the 4-body system. 
In addition, due to the $t$ = 1/2 isospin of the two fragments, the $^4$He excited states can either have isospin $T$ = 0 (isoscalar) or $T$ = 1 (isovector) quantum number. 
While in electron scattering all of these partial waves are involved, in $\alpha$ scattering only isoscalar $T=0$ states  with $S=0,1$ and $J=L$ can be populated. Therefore, besides the sought for $^{1}S_0$ resonance, only the $^{3}P_1$ partial wave located at 24.25 MeV contributes, making $\alpha$ scattering advantageous for our purposes.

Herein we report on a new study of the elastic and inelastic $\alpha$ + $\alpha$ scattering at 53 MeV incident energy. The 0$^+_2$ resonance and a small portion of the spectrum above the resonance energy are investigated while keeping low the number of open reaction channels.
In particular, the $\alpha$ + $\alpha$ $\rightarrow$ $\alpha$ + $p$ + $^3$H and the $\alpha$ + $\alpha$ $\rightarrow$ $\alpha$ + $n$ + $^3$He channels are experimentally accessed for the first time by exclusive experiments, making it possible to reconstruct the $^4$He excitation energy with unprecedented low background, high resolution and selectivity.
In this work we extract the energy and width of the $0^+_2$ state from a modern analysis of a new $\alpha$-$\alpha$ experiment (red point in Fig.~\ref{centroids_widths_b}), achieving the highly desired precision and providing a benchmark for future calculations.

While significant progress has recently been made in the \emph{ab initio} treatment of nuclear reactions~\cite{Elhatisari2015,elhatisari2025abinitiolatticestudy,Idini2019,Mercenne2023,Kravvaris2024,Rotureau2017}, no existing method can yet describe the inelastic $\alpha$--$\alpha$ scattering process as an eight-body problem in the continuum. In this work, we model the process as a two coupled-channel system composed of the ground state $0^+_1$ and the excited state $0^+_2$ (treated as a bound state),  with a  $0^+_1 \rightarrow 0^+_2$ coupling channel.
This approach is rooted in the well-established double-folding potential reaction theory.  However, instead of using phenomenological densities as done in the traditional theory, 
we employ here state-of-the-art \emph{ab initio} densities  that stem from different nuclear Hamiltonians, the same used in analyzing electron scattering data.
This represents a significant advance, and constitutes a solid framework for a first analysis of the $\alpha-\alpha$ scattering data presented here, that incorporates the microscopic physics presently available.
Our analysis of the line shape of the 0$^+_2$ state and of the elastic and inelastic differential cross sections, indicates that  densities from chiral EFT lead to a better agreement with experiment. However, space is left for an improvement of nuclear reaction models towards a full ab-initio description of the process.

\section{Results}
\subsection{Experimental details}\label{expdetails}
We performed an $\alpha$-scattering experiment at the K800 Superconducting Cyclotron of the Istituto Nazionale di Fisica Nucleare – Laboratori Nazionali del Sud (INFN-LNS) in Catania (Italy), using a beam energy of 53 MeV. Thin metallic films with implanted $^4$He atoms were used as targets. The $^4$He ejectiles emitted at forward angles were detected by the MAGNEX magnetic spectrometer~\cite{cap2016epja52} while the $^3$H and $^3$He fragments from the decays of the populated $^4$He excited state ($^4$He$^*$) were detected by the OSCAR array of solid state telescopes \cite{del2018nima877}. A schematic view of the set-up is given in Supplementary Fig. 1. The ejectiles were identified event by event in atomic number ($Z$), mass ($A$) and charge (see Methods \ref{Experiment} and Supplementary Fig. 2). The trajectories of the reaction ejectiles were then reconstructed and their momenta deduced at the target position (see Methods \ref{Methodsspe}). \par
This set-up allows on the one hand to characterize the $0^+_2$ resonance with unprecedented precision, and on the other hand, together with theoretical modeling, to provide stringent quality checks for ab-initio nuclear Hamiltonians.
 
\subsection{The resonance characterization}\label{resonance}

\begin{figure}[!]
\includegraphics[scale=0.33, trim = 0cm 0cm 0cm 13.7cm]{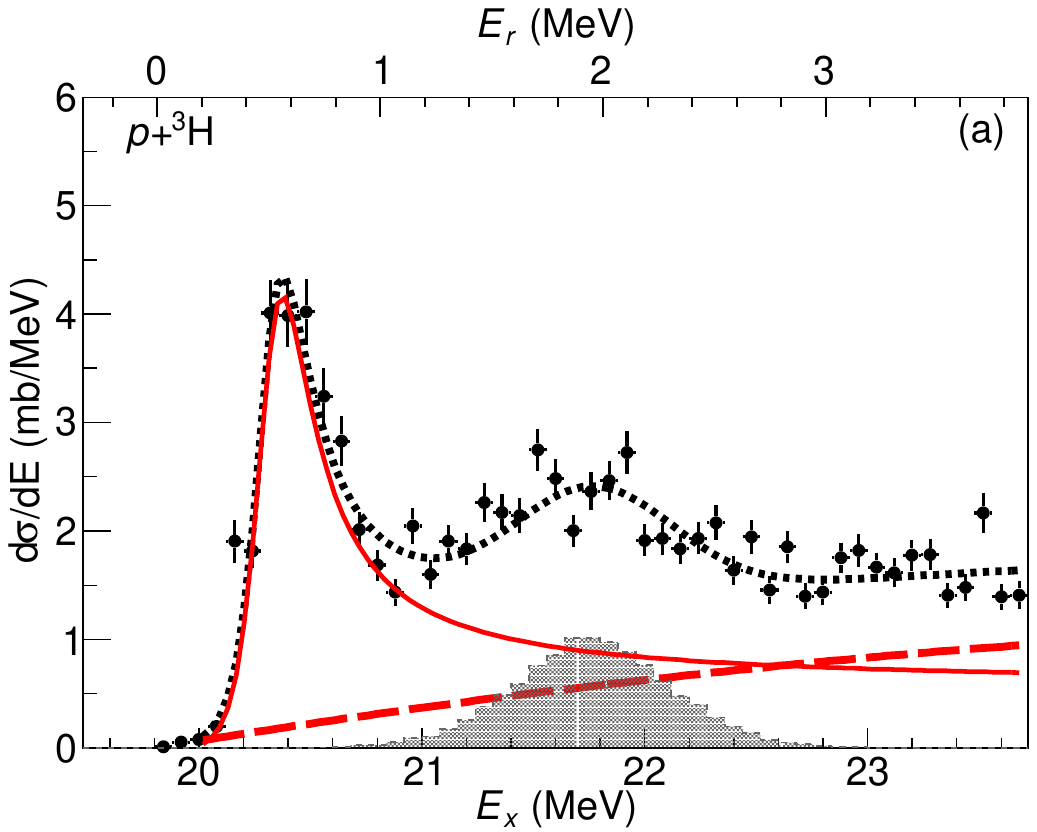}
\includegraphics[scale=0.33, trim = 0cm 0cm 0cm 13.7cm]{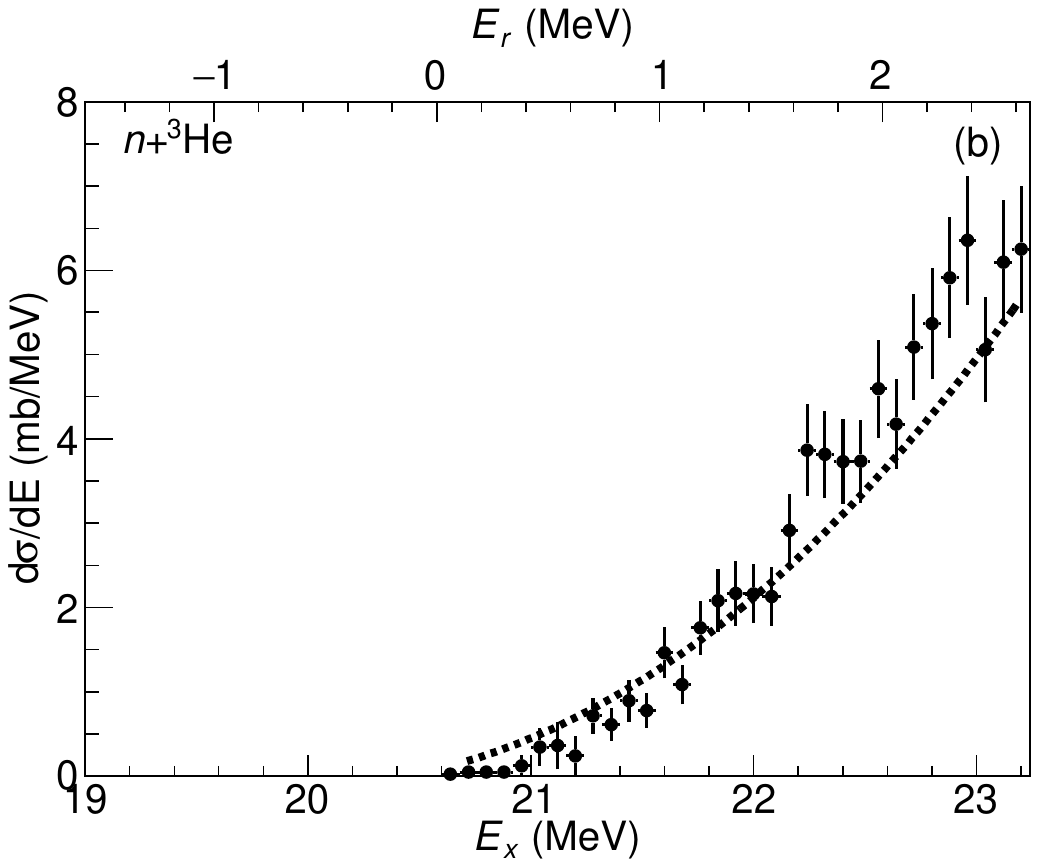}
\caption{\textbf{Energy spectra for the $p$ + $^3$H and $n$ + $^3$He channels.} (a) Experimental inelastic scattering spectrum (black dots) of the $p$ + $^3$H channel compared with the sum of three components (black dotted): $^{1}S_0$ (red solid),  $^{3}P_1$ resonance at 24.25 MeV ($E_r$ = 4.437 MeV) (red dashed) and $\alpha$ + $\alpha$ $\rightarrow$ $^3$H + $^5$Li $\rightarrow$ $\alpha$ + $p$ + $^3$H pick-up break-up transfer (grey hatched area). (b) The $n$ + $^3$He channel, dominated by the simulated (black dotted) $^{3}P_1$ resonance at 24.25 MeV ($E_r$ = 3.672 MeV). In both panels, error bars refer to the statistical and efficiency estimation uncertainties. The $E_r$ (upper horizontal axis in top) refers to the centroid-energy above the $S_p$ (a) and $S_n$ (b) threshold, respectively. \label{Fig_3hp_3hen_b}}
\end{figure}

In the inelastic process, the break-up channels  $p$ + $^3$H and  $n$ + $^3$He are experimentally accessed for the first time in hadronic experiments, making it possible to reconstruct the $^4$He excitation energy with very low background, high resolution and selectivity.
Fig.~\ref{Fig_3hp_3hen_b} shows the reconstructed $^4$He energy spectra from the  $p$ + $^3$H (Fig.~\ref{Fig_3hp_3hen_b}(a)) and  $n$ + $^3$He (Fig.~\ref{Fig_3hp_3hen_b}(b)) exit channels. In the $p$ + $^3$H the resonant $0_2^+$ state is clearly populated over the non-resonant background, presenting an overall signal to background ratio of $\approx$ 12.
We describe the   $p$ + $^3$H channel including: 
$(i)$ the  $^{1}S_0$ partial wave, featuring the resonant 0$^+_2$ state and the non – resonant breakup;
$(ii)$ the  $^{3}P_1$ component, through the 1$^-$ resonant state at 24.25 MeV ($E_r$ = 4.437 MeV) ~\cite{har2001};
$(iii)$ the pick-up breakup processes $\alpha$ + $\alpha$ $\rightarrow$ $^3$H + $^5$Li $\rightarrow$ $\alpha$ + $p$ + $^3$H centered at 21.78 MeV ($E_r$ = 1.967 MeV), as it emerges from the results of transfer reaction simulations. Although this structure could partially overlap with the $0_2^+$ state, we estimate an overall contribution of less than 0.2$\%$ in the region of the resonance between 19.72 and 20.84 MeV, which gives a negligible uncertainty in the $0_2^+$ cross section once we subtract its yield.

For the  $n$ + $^3$He channel we account for:
$(i)$ the  $^{1}S_0$ non – resonant breakup (the 0$^+_2$ resonance is below the threshold for this channel); 
$(ii)$ the  $^{3}P_1$ resonance at 24.25 MeV ($E_r$ = 3.672 MeV above the $S_n$ threshold);
$(iii)$ the $\alpha$ + $\alpha$ $\rightarrow$ $^3$He + $^5$He $\rightarrow$ $\alpha$ + $n$ + $^3$He pick-up break-up reaction.

In the $p$ + $^3$H channel, the interference of the 0$^+_2$ resonance with the $^{1}S_0$ non-resonant continuum and the presence of the tail of the $^{3}P_1$ partial wave clearly creates an asymmetry in the line shape.
A powerful way to study such an interference was originally  proposed by U. Fano \cite{fan1961pr124} and progressively extended in the realm of modern physics \cite{luk2010nat9,ott2013sci340,lim2017nat11,pal2021nat12}. Recently, it was used to characterize low-lying  resonances in the $^{10}$Li spectrum~\cite{cav2017prl118}, inspiring us to use it also  in the present study, leading to the very precise datum shown in Fig.~\ref{centroids_widths_b}.
Upon implementing the Fano analysis (see details in Methods, section \ref{Methods4}), $E_r$ is found at 0.50 $\pm$ 0.02 MeV (corresponding to an excitation energy on top of the ground state of $E_x$ = 20.31 $\pm$ 0.02 MeV). Including the uncertainties, our result is found within the overlapping range of values from previous studies from electron scattering (0.29 - 0.45) and hadronic scattering (0.47 - 0.48). 
Remarkably, the maximum of the observed peak is at $E_r$ = 0.59 ($E_x$ = 20.4 MeV), which is about 0.1 MeV higher than the actual resonance centroid energy as a result of distortion in the line shape due to the interference. 
The extracted width $\mathit{\Gamma_r}$ is 0.29 $\pm$ 0.04 MeV, in agreement with electron scattering studies but not with the $\alpha$ scattering analysis of Ref.~\cite{bau1981npa368}, reporting 0.89 MeV. 
Instead, assuming a symmetric Lorentzian shape $E_r= 0.59 \pm 0.04 $ and $\mathit{\Gamma_r}$ = 0.33 $\pm$ 0.05 are found. Thus, a systematic shift of about 90 keV for the resonance centroid and about 40 keV for the width is observed when the Lorentzian shape is adopted. 

\subsection{Angular-dependent cross section}\label{inelastic}
Here, we perform new calculations for the elastic and inelastic scattering in the framework of the coupled channel quantum scattering theory and compare them to our experimental data. 
The ISI and FSI are built in a semi-microscopic approach  by folding the nucleon-nucleon interaction with the 0$^+_1$ and 0$^+_2$ densities as well as the 0$^+_1$ $\rightarrow$ 0$^+_2$ transition density obtained from four different ab-initio calculations, all treating the $0^+_2$ resonance as a bound state. 
The obtained ISI and FSI were validated though comparisons with elastic scattering and reaction cross section data (see Methods, section \ref{Methodsuncert} for details). More details on the reaction modeling uncertainties can be found in section \ref{Methodsuncert} and Supplementary Figs. 5, 6. First, we consider densities from Bacca et al.~\cite{bac2013prl110,bac2015prc91} and Viviani et al.~\cite{viv2024fbs65}, which use a chiral EFT Hamiltonian based on a two-body force at next-to-next-to-next-to leading order~\cite{ent2003prc68}, and a different parametrization of the  three-body force at next-to-next-to-leading order \cite{nav2007fbs41,bar2018prc98}. We will display  cross sections obtained from these two sets of densities in one band encompassing the differences and name it ``the chiral EFT" family.
Second, we consider densities obtained from calculations with two  simplified Hamiltonians: one  from Hiyama et al.~\cite{hiy2004prc70},  where the three-body force is purely central, and one from Mei\ss{}ner et al.~\cite{mei2023arx}, where the Hamiltonian is based on the SU(4) isospin symmetry. 
We will display cross sections obtained from these two sets of densities as a band encompassing the differences and denote it as the ``simplified Hamiltonian" family.
These two bands serve as an uncertainty estimate quantifying the dependence of the cross section on the input densities.

\begin{figure}[!]
\includegraphics[scale=0.33, trim = 0cm 0cm 0cm 13.5cm]{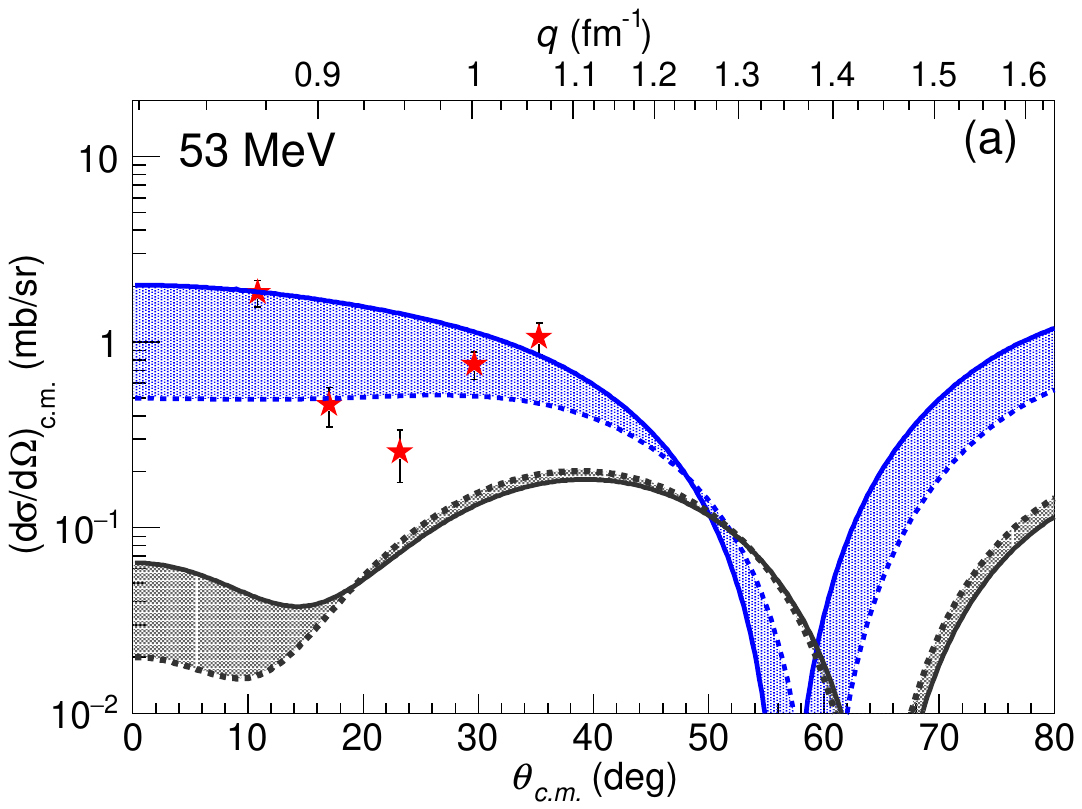}
\includegraphics[scale=0.33, trim = 0cm 0cm 0cm 13.5cm]{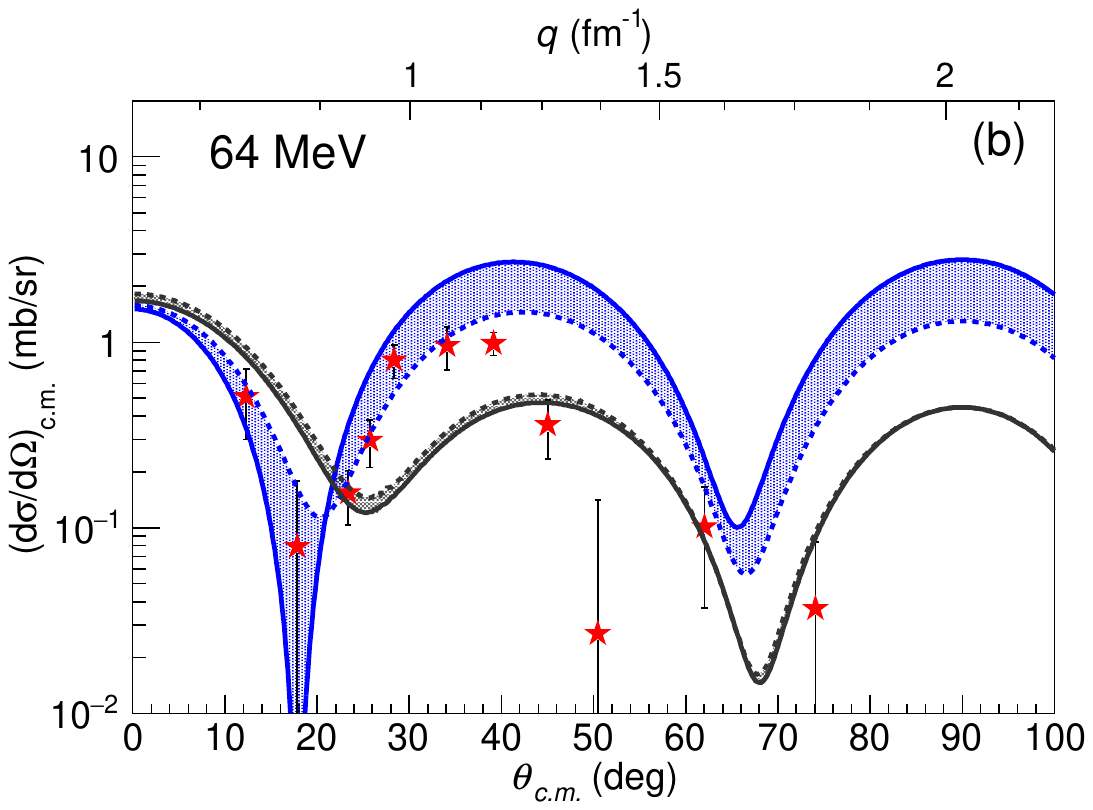}
\caption{ \textbf{Inelastic scattering angular distributions.} Inelastic scattering differential cross section for the 0$^+_2$ resonance as a function of the scattering angle in the centre--of--mass frame (lower horizontal axis) or the momentum transfer (upper horizontal axis) at 53 MeV (a) and 64 MeV \cite{gro1969pr178} (b) incident beam energy. The experimental data (red stars) are compared with the theoretical results obtained using the (blue band) chiral effective field theory densities (solid~\cite{bac2013prl110} and dashed~\cite{viv2024fbs65}) and the (grey band) simplified Hamiltonian densities (solid~\cite{mei2023arx} and dashed~\cite{hiy2004prc70}). The error bars correspond to the combination of uncertainties coming from the solid angle and detection efficiency estimation, the uncertainties in the beam current and the target thickness determination, the statistical error and the background subtraction. See text for details. \label{Fig_inel}}
\end{figure}

Fig.~\ref{Fig_inel} shows the experimental angular distributions of the differential cross section in comparison to theoretical calculations.
On panel (a) we have our experimental data obtained at a beam energy of 53 MeV, while on the panel (b) we also show previously published data at a beam energy of 64 MeV~\cite{gro1969pr178}.
In both cases, we consider only the contribution of the 0$^+_2$ resonance extracted, angle by angle, from the peaked spectral shape centered at 20.4 MeV, after a polynomial modeling of the background underneath. This accounts in average for the $^{3}P_1$ as well as the  $^{1}S_0$ non-resonant and interference contributions.

A fair agreement is found between the data and the calculations adopting the chiral EFT densities, both at 53 and 64 MeV. The pronounced minimum present in the data at 64 MeV at $\theta_{c.m.}\approx 18{\degree}$ (corresponding to $q\approx 0.85~{\rm fm}^{-1}$) is well reproduced. However, no clear sign of the minimum observed at 53 MeV at $\theta_{c.m.}\approx 23{\degree}$ is present in the calculations.
Differently from chiral EFT calculations, results from the simplified Hamiltonian family deviate from the data at 53 MeV as well as at small angles for the 64 MeV energy. At large angles for 64 MeV, all theories predict a minimum at $\theta_{c.m.}\approx 65{\degree}$ but the data are significantly less accurate there.

Confronting the experimental inelastic cross sections at 53 MeV and 64 MeV incident energy a marked difference appears, especially in the region at low momentum transfer. Indeed, due to the high excitation energy, the population of the 0$^+_2$ state also determines a sudden change of the kinetic energy and wavelength for the colliding systems. This unveils an interesting situation where a small change in the projectile energy generates a substantial change in the diffraction pattern, paving the way for a strong sensitivity of this process to fine details of the scattering dynamics, namely the spatial distributions of the $^4$He $0^+_1$ and $0^+_2$ states.  
The pronounced first minimum observed at $\approx 0.85~{\rm fm}^{-1}$ for the cross section at 64 MeV energy is damped at 53 MeV and shifted at $\approx 0.95~{\rm fm}^{-1}$, while it almost disappears in the calculations. 

An expansion of the calculated inelastic scattering cross sections in terms of the angular momentum $l$ of the relative motion of the outgoing nuclei allows to get insight on this situation. At 64 MeV the cross section is mainly fed by the $l = 4$ partial wave, with minor role for the $l=0$ and $l=2$ and negligible contribution for $l \ge 6$. The $l=4$ partial wave features a pronounced minimum at about 20$\degree$, common to all calculations as well as to the data. Instead, the $l = 2$ contribution is the leading one at 53 MeV, also due to the less available momentum for the scattering waves, with a minimum at about 55$\degree$ and a growing trend down to zero degree. In the case of the simplified Hamiltonian, the same analysis still indicates a similar behavior at both the energies, although the $l=4$ contribution at 53 MeV is more relevant and a minimum starts to appear at forward angles ($\approx 15\degree$). 
As a consequence, the data at 53 MeV suggest the need for a mild redistribution of the cross section partial waves toward an enhanced $l=4$ contribution. In terms of radial distribution of the cross section this shift toward larger $l$ reflects the need of more density overlap of the two colliding nuclei at larger distances ($R\gtrsim 4~{\rm fm}$).
This is significantly larger than the size of $^4$He ground state radius ($r(^4$He$_{g.s.}) = 1.7~{\rm fm}$  \cite{wan2024prc109}), typically explored in electron scattering experiments indicating that with $\alpha$ scattering we are probing more peripheral regions of the density.
In particular, while the latter is sensitive only to the transition density, 
$\alpha$--$\alpha$ scattering is particularly sensitive to the density of the  $0^+_2$ excited state, for which the prediction of the four different ab-initio calculations differ substantially. 

We find that the best result is given by the chiral EFT density by Viviani et al.~\cite{viv2024fbs65} and prove that inelastic $\alpha$--$\alpha$ scattering poses stringent tests to nuclear structure models,
emphasized by the strong sensitivity of the diffractive patterns in the cross section angular distributions to the details of the resonances density profile.

\section{\label{sec:discussion}Discussion}

Our study allows to give insight on the nature of the 0$^+_2$ resonance of $^4$He and provides strong constraints for present and future models of the nuclear Hamiltonian. The adopted experimental technique allowed to get an unprecedented clean and background-free spectral response. Our results within the Fano theory of resonances give a centroid energy of $E_r$ = 0.50 $\pm$ 0.02 MeV and a width $\mathit{\Gamma_r}$ = 0.29 $\pm$ 0.04 MeV, providing the most precise ever measurements of the 0$^+_2$ resonance, in disagreement with previous $\alpha$-scattering data~\cite{gro1969pr178,bau1981npa368} but in agreement, as for the width, with the latest electron scattering experiment by Kegel et al.~\cite{keg2023prl130}. Since the Fano analysis indicates a significant interference between the 0$^+_2$ resonance and the underlying non resonant continuum, employing a similar analysis on other scattering data would be interesting.

The comparisons of the experimental cross sections with calculations based on coupled channels quantum scattering theory, with ab-initio densities used in input, has allowed to set a robust test of $^4$He wave functions of the ground state 0$^+_1$, the 0$^+_2$ resonance, and the 0$^+_1$ $\rightarrow$ 0$^+_2$  transition density. Interestingly, we find that densities from chiral EFT 
better describe the $\alpha$ scattering data. Due to the importance of the 0$^+_2$ density in the $\alpha-\alpha$ scattering calculation, we conclude that while the densities from Refs.\cite{hiy2004prc70,mei2023arx} well reproduce the transition form factor probed by electron scattering, they do not well describe the density of the excited state probed by the $\alpha$ scattering. The chiral densities of Viviani et al.~\cite{viv2024fbs65} provide the best overall agreement with data from both electron and $\alpha$ scattering, despite a slight deviation from the diffraction pattern observed in $\alpha$ scattering at 53 MeV. \par

At the same time, the fact that the interaction used in Ref.~\cite{viv2024fbs65} predicts a resonance position and width that deviate from the experimentally extracted values (see Fig. \ref{centroids_widths_b}) highlights a deeper challenge. Indeed, establishing a transparent link between the measured cross section in an open quantum system, such as $\alpha$–$\alpha$ scattering and the underlying resonance in one of its sub-systems, remains a challenging problem in quantum mechanics.
Despite notable advances in reaction theory, a fully quantitative framework applicable to inelastic $\alpha$–$\alpha$ scattering is still lacking. Our analysis—based on a robust yet necessarily approximate and not fully consistent reaction model—demonstrates that combining $\alpha$-scattering with electron-scattering  substantially tightens the sensitivity to the underlying microscopic physics. These results not only expose the current theoretical limitations but also clearly define the direction for future progress, including the long-term goal of constructing an ab-initio optical potential for light nuclei. Ultimately, the high-precision data reported here, together with the insights gained from this study, provide a compelling impetus for advancing theoretical approaches to inelastic $\alpha$–$\alpha$ scattering.


\section{Methods}\label{Methods}
\subsection{Experimental set-up and particle identification}\label{Experiment}

Two experiments were performed at the MAGNEX facility \cite{cap2016epja52} of the Istituto Nazionale di Fisica Nucleare – Laboratori Nazionali del Sud (INFN-LNS) in Catania.\par 
The first experiment was dedicated to the elastic scattering measurement in the angular range 6.5$\degree$ $\le$ $\theta_{\rm Lab}$ $\le$ 45$\degree$, corresponding to 13$\degree$ $\le$ $\theta_{\rm c.m.}$ $\le$ 90$\degree$ in the centre of mass reference frame. The $^4$He beam was accelerated by the K800 Superconducting
Cyclotron at 53 MeV (13.2 MeV/u) incident energy and impinged on a solid target, built by the implantation of $^4$He atoms on a $^{181}$Ta sheet. The effective thickness of $^4$He component of the target, measured by Rutherford back-scattering (RBS) technique, was 2.5 $\pm$ 0.13 $\mu$g cm$^{-2}$  
(3.7x10$^{17}$ atoms cm$^{-2}$) while the thickness of $^{181}$Ta was 950 $\pm$ 10 $\mu$g cm$^{-2}$ (3.2x10$^{18}$ atoms cm$^{-2}$). Residual impurities of oxygen (14.6 $\pm$ 0.4 $\mu$g cm$^{-2}$ (5.5x10$^{17}$ atoms cm$^{-2}$)) were observed in the RBS analyses. The beam current was measured by a Faraday cup (FC) mounted 15 cm downstream of the target. The solid angle of MAGNEX was 50 msr, defined by four 1 mm thick tantalum slits located 25 cm downstream of the target. The $\alpha$-particle ejectiles were detected by the MAGNEX Focal Plane Detector (FPD) \cite{tor2021nima989}, which consists of a gas-filled tracker followed by a wall of 60 (5 x 7 cm$^2$ area, 1 mm thick) silicon pad detectors.\par

In the second experiment, conducted at the same facility, the inelastic scattering was studied by performing a coincidence measurement between the $^4$He ejectiles and one of the breakup fragments emitted by the residual $^4$He ($^3$H or $^3$He). Our study was designed such as to explore a significant portion of the phase space for both the $^3$H + $p$ and $^3$He + $n$ decay modes. 
The experimental set-up is illustrated in Supplementary Fig. 1. Additional elastic scattering data were collected under experimental conditions similar to the first experiment to cross check.
The $^4$He beam was accelerated at 53 MeV incident energy and impinged on a solid target built by the implantation of $^4$He atoms on a $^{27}$Al sheet. 
The replacement of the tantalum backing with an aluminium one was intended to further reduce the background observed in the first experiment, attributed to beam scattering on tantalum. 
The effective thickness of $^4$He was 1.3 $\pm$ 0.1 $\mu$g cm$^{-2}$ (1.9x10$^{17}$ atoms cm$^{-2}$) while the thickness of $^{27}$Al was 210 $\pm$ 2 $\mu$g cm$^{-2}$ (4.6x10$^{18}$ atoms cm$^{-2}$). Residual impurities of oxygen (2.7 $\pm$ 0.1 $\mu$g cm$^{-2}$ (1.0x10$^{17}$ atoms cm$^{-2}$)) and carbon (0.2 $\pm$ 0.02 $\mu$g cm$^{-2}$ (1.2x10$^{16}$ atoms cm$^{-2}$)) were observed in the RBS analyses. 
The beam current was measured by a FC mounted 15 cm downstream of the target. The measurement was repeated with an ancillary target composed of aluminium, oxygen and carbon in the appropriate proportions for background estimation. 
The $^4$He ejectiles were momentum analysed by the MAGNEX spectrometer, whose optical axis was set at $\theta_{\rm opt}$ = 6.6$\degree$, spanning the range of detection angles between 2$\degree$ $\leq$ $\theta_{\rm Lab}$ $\leq$ 12$\degree$ in the laboratory reference frame. The momentum acceptance of the spectrometer ($\approx$ 24$\%$) \cite{cap2016epja52} allowed the measurement of the $^4$He excitation energy ($E_x$) up to 4.0 MeV above the $S_p$ threshold, covering a significant part of the available energy spectrum, which extends up to 6.7 MeV above $S_p$ at the beam energy of the experiment. The $\alpha$-particle ejectiles were detected by the MAGNEX FPD, while the $^3$H and $^3$He fragments were detected by the OSCAR solid state telescope (5 x 5 cm$^2$ area) \cite{del2018nima877}. This device consists of two detection stages: a Single Sided Silicon Strip Detector (SSSSD), which measures the energy loss $\Delta$E, followed by 16 Silicon pads (arranged in a 4 x 4 mode) for measuring the ions residual energy $E_{resid}$. The thicknesses of 20 $\mu$m for the $\Delta$E and 300 $\mu$m for the $E_{resid}$ allow the unambiguous identification of light reaction fragments with energies of few MeV. In particular, the energy range of $^3$H and $^3$He particles detected by OSCAR was 2.5 - 9.0 MeV and 5.0 - 11.2 MeV, respectively. For both of them a large part of their phase space was explored. The OSCAR system was mounted inside the MAGNEX scattering chamber, 15 cm far from the target, subtending an angular range between 19$\degree$ and 38$\degree$ with an angular resolution of $\sim$ 1$\degree$.
The unambiguous particle identification is illustrated in Supplementary Fig. 2. Details on the particle identification technique can be found at Refs. \cite{cap2010nima621,del2018nima877}.\par

\subsection{Details of data reduction}\label{Methodsredu}

\subsubsection{Energy spectra}\label{Methodsspe}

The trajectories of the $^4$He ejectiles detected at the MAGNEX FPD were reconstructed and the momentum vector was deduced at the target position, adopting the methodology described in Ref. \cite{cap2011nima638}. The missing mass technique, with the use of relativistic kinematics relations, was adopted to get $^4$He energy spectra.
The inelastic scattering events were further characterized by the coincident measurement of $^3$H or $^3$He fragments by the OSCAR telescope (see section \ref{Methods3}). The obtained energy resolution, in the region of the $^4$He(0$^+_2$) state, ranges from $\delta_E \approx$ 80 keV at small scattering angles to $\delta_E \approx$ 100 keV for the largest explored angles. This was mainly determined by the spectrometer response, the beam energy spreading and the kinematic energy dependence on the scattering angle. A weighted energy resolution of 86 keV was estimated when summing the spectra of all explored angles.\par

\subsubsection{Line shape analysis}\label{Methods4}
An important aspect in analyzing the $^4$He(0$^+_2$) resonance is to include the contribution of the $L > 0$ resonant and non-resonant strengths and the possible interference between the 0$^+_2$ resonance with the underlying non-resonant continuum featuring the same 0$^+$ quantum numbers. A powerful way to study this topic was proposed by Fano \cite{fan1961pr124}, stemming from the asymmetric line shape (modeled by the Fano function) generated by such interference. 
\par
An additional feature is the distortion of the resonance line shape due to the proximity of the particle emission threshold \cite{mcv1994npa576}. Since the strength is zero below the threshold an asymmetric line shape is always found, although the actual shape depends also on the multipolarity. Differently from
the Fano asymmetry, the threshold effect does not depend on the dynamic mechanism populating the resonance, thus providing a stable distortion of its line shape. The stronger is the distortion the larger is $\mathit{\Delta_r}$ = $\mathit{\Gamma_r}$ /($E_r$ - $E_t$), the ratio between the resonance full width at half maximum ($\mathit{\Gamma_r}$) and the distance of the resonance centre ($E_r$) from the threshold ($E_t$), especially when $\mathit{\Delta_r}$ exceeds unity. In the $p$ + $^3$H channel the threshold at $S_p$ = 19.813 MeV can affect the line shape for both the $^1S_0$ ($\mathit{\Delta_r}$ $\approx$ 0.7) and the
isoscalar $^3P_1$ ($\mathit{\Delta_r}$ $\approx$  1.4) resonances, with a smaller impact on the former thanks to its smaller width. Ideally, both the interference with non-resonant continuum and the threshold effect should be included in the resonance line shape analysis. However, to our knowledge such a theoretical framework is not available to date and approximate schemes are needed.\par
Our assumption for the $^{1}S_0$ resonance is to account only for the Fano asymmetry, since $\mathit{\Delta_r} < 1$. In reverse, we assume only the threshold effect for the isoscalar $^{3}P_1$, considering the small $L = 1$ non-resonant strength in the explored excitation energy region \cite{viv2020prc102}. 
Fig. \ref{Fig_3hp_3hen_b} (a) shows the best fit obtained adopting a Fano function, convoluted with experimental energy resolution, as a model of the observed structure in the region of the $0_2^+$ state. In this framework the cross section in the region of the resonance is given as 
\begin{equation}
\sigma = \sigma_{cont}\frac{\left|q_F+\epsilon\right|^2}{1+\epsilon^2} , 
\label{eqfano} 
\end{equation}
where $\sigma_{cont}$ is the inelastic cross section of the non-resonant continuum and $\epsilon$ = $2(E-E_r)/\mathit{\Gamma_r}$, 
where $E_r$ is the resonance centre and  $\mathit{\Gamma_r}$ its width. 
The line shape is controlled by the Fano parameter $q_F$ which, apart from a phase factor, is determined by the relative weight of the resonant and non-resonant amplitudes. An interesting aspect of the Fano theory is that the $q_F$ parameter depends on the dynamics of the resonance population, while the resonance centre and width do not change. When $q_F \gg 1$  the line shape becomes progressively more symmetric approaching the Lorentzian function when $q_F$ goes to infinite.
For this analysis, we find $q_F$ = 2.54 $\pm$ 0.07 when we consider the spectrum in the full explored angular region, which calls for a significant effect from the interference discussed here. Further analyses of the line shape in smaller angular slices of the measured cross section do reveal that the resonance centre and width are stable, within the quoted uncertainties.\par
The quoted uncertainties for the resonance energy ($\pm$0.02 MeV) and width ($\pm$0.04 MeV) include the uncertainty in the energy threshold effect on the line shape model, the deconvolution of the spectrum in the three components (as shown in Fig. \ref{Fig_3hp_3hen_b}), the contribution of the finite bin size, the accuracies in energy or/and angle of the delivered beam, the spectrometer and the ray-reconstruction procedure as well as the variation of the experimental resolution as a function of angle (see Methods (\ref{Methodsspe})).\par

\subsubsection{Cross section determination}\label{Methods3}

The elastic scattering yields were determined with angular steps of 0.5$\degree$ in MAGNEX over the explored angular range. The differential cross-section was extracted through a careful determination of the experimental yield, the beam flux, the target thickness, the explored solid angle and the detection efficiency. For the inelastic scattering process, exclusive yields were determined with angular steps of 2$\degree$ in MAGNEX, combined with the whole OSCAR system, to reach a compromise with the reduced yield of the coincidence measurement. 
The background due to aluminium, oxygen and carbon impurities in the target as well as that generated by spurious signals in the detectors was negligible in the coincidence spectra, leading to an overall signal to instrumental background ratio ranging between 15 and 40 at the different detection angles. Nonetheless, this background was experimentally determined using data collected with the ancillary target and subtracted.

The efficiency of our detection system in the coincidence measurement was calculated by the MULTIP Monte Carlo simulation algorithm \cite{sgo2017epja53}, taking into account the MAGNEX intrinsic efficiency \cite{cav2011nima637}. The exclusive yields, corrected for the detection efficiency, were transformed to laboratory double differential cross sections taking into account the areal density of the $^4$He scattering centres, the beam current, the MAGNEX and OSCAR solid angles, the last one computed by a Monte Carlo simulation.\par

\subsection{Calculation methodology}\label{Methodstheo}

The theoretical cross section angular distributions were calculated in the coupled - channel quantum scattering theory framework. The nucleus - nucleus potential was obtained by double folding the nucleon - nucleon interaction with densities from four different ab-initio theories. 
A first group consisted by two theories of the chiral EFT family by Bacca et al.~\cite{bac2013prl110,bac2015prc91} and Viviani et al.~\cite{viv2024fbs65} while, a second one, by theories on which the densities were obtained from calculations with two simplified Hamiltonians by Hiyama et al.~\cite{hiy2004prc70} and by Mei\ss{}ner et al.~\cite{mei2023arx}. The $\rho_{0^+_1}$ and $\rho_{0^+_2}$ densities for the $0^+_1$ and $0^+_2$ state respectively, as well the transition density $\rho_{0^+_1\rightarrow 0^+_2}$ are presented in Supplementary Fig. 3(a-c). One can find more details on the reaction theory, the double folding model and the densities in Supplementary Methods (1.1), on the adopted potentials in the reaction theory in Supplementary Methods (1.1.1) and on the used ab-initio theory in Supplementary Methods (1.2). Finally, in Supplementary Methods (1.2.1), one may see how we successfully handle the $0^+_2$ lying in the continuum as an effective bound state. The reliability of the process is illustrated in Supplementary Fig. 3(d). The calculated centroid-energy, assuming the $0^+_2$ as an effective bound state, is $E_r$ = 0.5 $\pm$ 0.1 MeV. \par

\subsection{Assessment of the reaction modeling uncertainties}\label{Methodsuncert}

As for the uncertainties in the reaction modeling, the main source comes from limitations in the adopted model space and uncertainties in the determination of the ISI and FSI. Indeed, our coupled - channel reaction scheme includes explicitly only the elastic scattering and the inelastic excitation of the 0$^+_2$ resonance. Other inelastic channels, although kinematically accessible, are accounted for in average through a complex polarization nucleus - nucleus potential. A phase - shift analysis with the adopted potentials was performed as described in Supplementary Methods (1.1.1), whose results are illustrated in Supplementary Fig. 4. The availability of a rich and accurate dataset at $^4$He + $^4$He interaction at different energies and for different partial waves \cite{rus1956pr104,hey1956pr104,nil1956pr104,bac1972prl29,tom1963pr129,chi1974prc10,cha2011prc83,bre1959prsl251,dar1965phd,dar1965pr137} gave us the chance to fine tune the potential and evaluate a realistic uncertainty band. In particular, the range of the band was set in order to allow variations of the $|S|$ matrix within an interval $\Delta|S|$ = $\pm$ 0.1 around the best value. The goodness of this approach is first evaluated through comparisons of the calculated elastic scattering angular distribution cross sections at $E_{Lab} = 53$ MeV with our data as shown in Supplementary Fig. 5 (a), presenting an overall good agreement. The existing elastic scattering data at 53.4 MeV from Ref.~\cite{dar1965pr137} are also in good agreement with the new dataset. Our theoretical framework has been adopted to analyze also the existing elastic scattering data at $E_{Lab} = 64$ MeV from Ref.~\cite{dar1965pr137}. Remarkably, the theoretical results are in very good agreement with these data as visible in Supplementary Fig. 5 (b). Overall, the results point to a strong reliability of our phase shift analysis and consequently of our ISI. Further comparisons were performed between the total reaction cross section ($\sigma_{\rm tot}^{\rm theo}$), from coupled-channel calculations with that extracted from suitable experimental data and reliable fusion reaction calculations ($\sigma_{\rm tot}^{\rm exp}$) \cite{gut1973npa213,gav1980prc21,tar2008nimb266}. The $\sigma_{\rm tot}$ observable is notoriously related to the imaginary part of the ISI. The result of this comparison is presented in Supplementary Table 1, presenting a very good agreement between experimental ($\sigma_{\rm tot}^{\rm exp}$) and theoretical ($\sigma_{\rm tot}^{\rm theo}$) reaction cross sections, although the error bars are sizable in both cases. The effect of the nucleon - nucleon interaction was found negligible by changing the adopted Brazilian Nuclear Potential (BNP) with the Hamiltonians used in the ab-initio densities. \par
Finally, the effect of the uncertainty band in the $^4$He + $^4$He interaction in the inelastic angular distribution cross sections is presented in Supplementary Fig. 6. \par


\bmhead{Data availability}
The data are available from the corresponding author upon reasonable request.



\bmhead{Acknowledgments}
We warmly acknowledge Prof. H. Lenske, Prof. N. Barnea, Prof. W. Leidemann and the late Prof. A. Vitturi for enlightening discussions. We warmly acknowledge Prof. E. Hiyama, Prof. U.-G Mei\ss{}ner and Dr. S. Shen for providing us with their density distributions. We warmly acknowledge the technical staff of INFN-LNS for the support. We also warmly acknowledge M. D' Andrea for the technical support with the OSCAR set-up. The project has received funding from the European Research Council (ERC) under the European Union's Horizon 2020 research and innovation programme (grant aggreemnt No. 714625). S.B. acknowledges support by the Deutsche Forschungsgemeinschaft (DFG) through the Cluster of Excellence ``Precision Physics, Fundamental Interactions, and Structure of Matter'' PRISMA${}^+$ EXC 2118/1 (Project ID 39083149) and through CRC1660: Hadrons and Nuclei as discovery tools (Project No. 514321794). L.C. acknowledges financial support by Funda\c{c}\~ao de Amparo \`a Pesquisa do Estado de S\~ao Paulo (FAPESP) Proc. No 2019/07767-1, Conselho Nacional de Desenvolvimento Cient\'ifico e Tecnol\'ogico (CNPq) Proc. No 302072/2022-5, and project INCT-FNA Proc. No 464898/2014-5. 

\bmhead{Author Contributions}
F.C., S.B., D.C., M.Cav., C.A., G.O and L.C. proposed the experiment.
F.C., V.S., D.C., M.Cav. and I.L. performed the experimental set-up with a contribution from H.-W.B., S.C., C.C., M.Cin., O.S., A.S. and M.Vig..
F.C., V.S., D.C., M.Cav., I.L., C.A., G.A.B., S.C., M.Cic., I.C., C.F., A.H., M.H., Y.K., O.S., A.S., D.T. and A.Y. performed to the data taking. 
V.S. performed the data reduction with a contribution from F.C., D.C., M.Cav., M.F., M.Cic., M.Cin., D.D., A.H., T.M. and O.S.. 
V.S. and F.C. performed the line shape analysis. 
L.C. performed coupled-channel cross section calculations with a contribution from F.C., V.S. and Y.K.. 
S.B., G.O., M.Viv., A.K. collected or/and calculated densities from ab-initio calculations. 
F.C., V.S., S.B., L.C., G.O. and M.Viv. wrote the manuscript. All the authors have revised the manuscript.

\bmhead{Competing interests}
The authors declare no competing interests.

\bmhead{Supplementary Information}
Supplementary Information accompanies this paper.


\pagebreak 

\includepdf[pages=-]{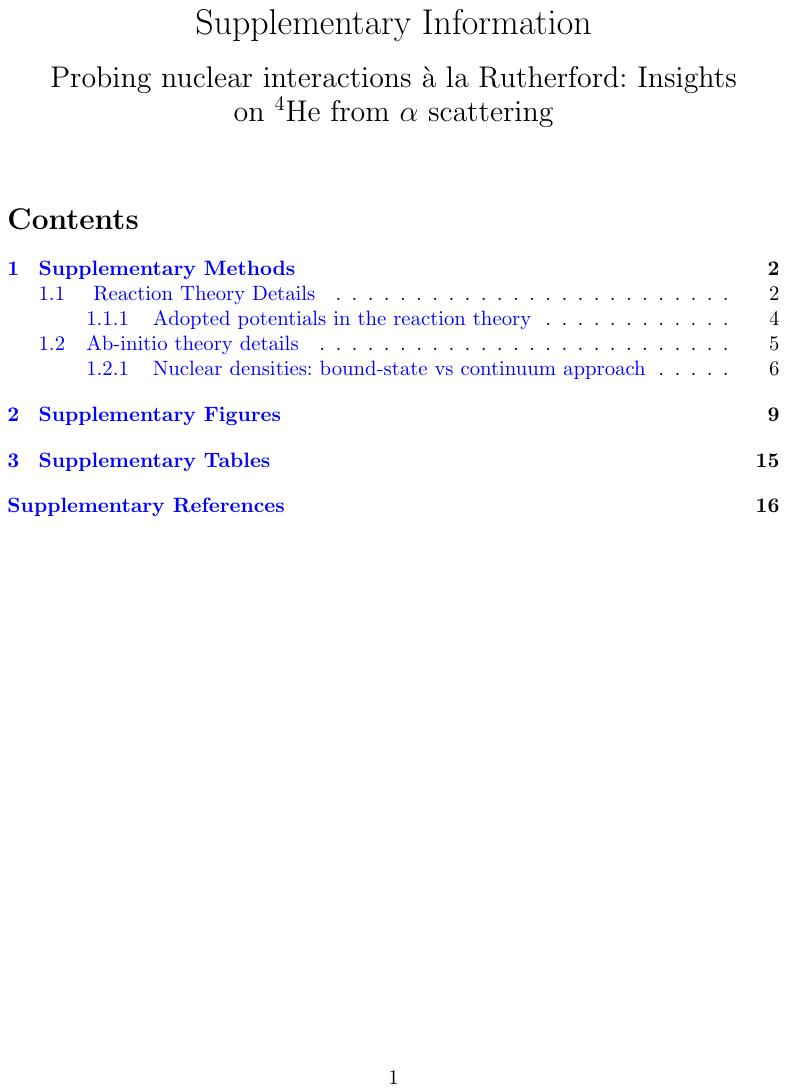}

\end{document}